\documentclass[sigconf]{acmart}
\AtBeginDocument{%
  }

\setcopyright{none}
\acmConference[UIST '26]{...}
\acmYear{2026}

\usepackage{mdframed}
\usepackage{listings}
\usepackage{todonotes}
\usepackage{microtype}

\usepackage[most]{tcolorbox}
\usepackage{xcolor}
\usepackage{listings}

\definecolor{promptbg}{RGB}{250,250,250}
\definecolor{promptframe}{RGB}{185,185,185}
\definecolor{prompttitlebg}{RGB}{242,242,242}

\definecolor{systemcolor}{RGB}{44,92,152}
\definecolor{usercolor}{RGB}{156,82,35}
\definecolor{keywordcolor}{RGB}{120,40,140}

\newcommand{\systemName}{\textsc{YT-Pilot}}
\newcommand{\syskw}[1]{\textcolor{systemcolor}{\textbf{#1}}}
\newcommand{\userkw}[1]{\textcolor{usercolor}{\textbf{#1}}}
\newcommand{\keykw}[1]{\textcolor{keywordcolor}{#1}}

\newtcolorbox{promptbox}[1]{
  enhanced,
  breakable,
  colback=promptbg,
  colframe=promptframe,
  boxrule=0.45pt,
  arc=1.2mm,
  left=1.2mm,
  right=1.2mm,
  top=0.8mm,
  bottom=0.8mm,
  fonttitle=\bfseries\small,
  title={#1},
  colbacktitle=prompttitlebg,
  coltitle=black,
  boxed title style={
    colframe=promptframe,
    colback=prompttitlebg,
    boxrule=0.45pt,
    arc=1mm
  },
  attach boxed title to top left={xshift=1.5mm,yshift=-1mm},
  fontupper=\ttfamily\footnotesize
}

\NewDocumentCommand{\cheng}{ mO{} }{\textcolor{purple}{\textsuperscript{\textit{Cheng}}\textsf{\textbf{\small[#1]}}}}

\begin{document}

\title{\systemName: Turning YouTube into Structured Learning Pathways with Context-Aware AI Support}

\author{Dina Albassam}
\affiliation{
  \institution{University of Illinois Urbana-Champaign}
  \department{Computer Science}
  \city{Champaign}
  \state{Illinois}
  \country{USA}
}
\email{dinasa2@illinois.edu}

\author{Kexin Quan}
\authornote{These authors contributed equally}
\affiliation{
  \institution{University of Illinois Urbana-Champaign}
  \department{School of Information Sciences}
  \city{Champaign}
  \state{Illinois}
  \country{USA}
}
\email{kq4@illinois.edu}

\author{Mengke Wu}
\authornotemark[1]
\affiliation{
  \institution{University of Illinois Urbana-Champaign}
  \department{School of Information Sciences}
  \city{Champaign}
  \state{Illinois}
  \country{USA}
}
\email{mengkew2@illinois.edu}

\author{Sanika Pande}
\affiliation{
  \institution{University of Illinois Urbana-Champaign}
  \department{SALT Lab}
  \city{Champaign}
  \state{Illinois}
  \country{USA}
}
\email{sanikap2@illinois.edu}

\author{ChengXiang Zhai}
\affiliation{
  \institution{University of Illinois Urbana-Champaign}
  \department{Computer Science}
  \city{Urbana}
  \state{Illinois}
  \country{USA}
}

\author{Yun Huang}
\affiliation{
  \institution{University of Illinois Urbana-Champaign}
  \department{School of Information Sciences}
  \city{Champaign}
  \state{Illinois}
  \country{USA}
}

\renewcommand{\shortauthors}{Albassam et al.}


\begin{abstract}
YouTube is widely used for informal learning, where learners explore lectures and tutorials without a predefined curriculum. However, learning across videos remains fragmented: learners must decide what to watch, how videos relate, and how knowledge builds. Existing tools provide partial support but treat planning and learning as separate activities, lacking a persistent interaction structure that connects them. Grounded in self-regulated learning theory (SRLT), we introduce \systemName, a pathway-aware learning system that operationalizes the \textit{learning pathway} as a persistent, user-facing interaction structure spanning planning and learning. The pathway coordinates goal setting, planning, navigation, progress tracking, and cross-video assistance. Through a within-subjects study ($N=20$), we show that \systemName\ significantly improves perceived goal clarity, pathway coherence, and progress tracking, while shifting interaction toward pathway-level reasoning across multiple resources.
\end{abstract}




\maketitle

\section{Introduction}

YouTube has become one of the most widely used platforms for informal learning, where learners explore lectures, tutorials, and walkthroughs without a predefined curriculum \cite{kim2013,mansour2016,lange2019}. In such settings, learning is largely self-directed, requiring learners to set goals, plan content, monitor progress, and reflect over time, processes central to self-regulated learning theory (SRLT) \cite{zimmerman2002}. However, these processes remain weakly supported in practice: learners must decide what to learn, how to sequence content, and how knowledge builds across videos. While playlists can provide creator-defined sequences, they are typically static and non-personalized, and do not provide a persistent, user-facing interaction structure that supports goal setting, planning, and progress tracking across the learning process. As a result, learning across videos remains fragmented and requires learners to reconstruct their own learning process over time \cite{tan2013,june2014}.

Recommendation systems and YouTube’s Learning channel~\cite{youtubelearningchannel} provide partial support through grouped content and single-video assistance (see Appendix~\ref{fig:youtube}), while recent LLM-based tools and planning systems generate structured study plans or provide contextual assistance \cite{chun2025,wang2025learnmate,li2024tutorly,ge2025srlagent}. However, these systems remain fragmented: plans are ephemeral, not carried into learning, and interactions are not personalized or grounded in a persistent, user-facing interaction structure that supports self-regulated learning across phases. As a result, learners must continuously reconstruct goals, progression, and context across sessions.

To address this gap, we introduce the learning pathway as an explicit interaction structure grounded in SRLT that persists across planning and learning. Rather than treating plans, recommendations, or conversations as outputs, we operationalize the pathway as a persistent, user-facing structure through which learners set goals, construct conceptual roadmaps, navigate content, and track progress over time.

We instantiate this interaction paradigm through \systemName, a pathway-aware system that operationalizes the pathway across both planning and learning phases. During planning, the pathway supports goal setting, planning, and concept roadmap construction, while during learning it coordinates navigation, progress tracking, context-based note-taking, and pathway-aware assistant, enabling users to reason over learning as a structured trajectory rather than isolated videos or ephemeral plans.

We evaluate \systemName\ through a within-subjects study ($N=20$) comparing it to YouTube’s Learning channel across three research questions:

\textbf{RQ1 (Planning Phase):} How does a guided interface with preference-based goal setting and conceptual roadmap visualization support learners in shaping and configuring a learning pathway?

\textbf{RQ2 (Pathway Generation):} How does a structured video pathway influence learners’ ability to interpret, follow, and revise learning progression across videos?

\textbf{RQ3 (Learning Phase):} How does an integrated environment with progress tracking, pathway-aware assistant and context-based note-taking support learners in navigating, maintaining, and reasoning across a learning pathway during a session?

We make two contributions:

\textbf{(1) SRLT-grounded learning pathway as an explicit interaction structure.}
We introduce the learning pathway as an explicit, persistent interaction structure grounded in SRLT that spans planning and learning. Rather than treating plans as static outputs, we operationalize the pathway as the primary unit of interaction through which learners set goals, interpret progression, maintain learning, and reason across the pathway.

\textbf{(2) Empirical characterization of pathway-based interaction.}
Through a within-subjects study, we show that pathway-based interaction improves perceived goal clarity, pathway coherence, and progress tracking, while supporting cross-video reasoning and navigation.

\section{Related Work}

\subsection{Self-Regulated Learning in Informal Digital Environments}

Self-Regulated Learning (SRL) describes how learners actively manage learning through goal setting, monitoring, and reflection \cite{zimmerman2002}. While formal environments provide structured scaffolding, informal digital environments shift this responsibility to learners \cite{blaschke2012,wang2025oill}.
Prior work has explored supporting SRL through intelligent systems, including pedagogical agents \cite{azevedo2022metatutor}, LLM-based metacognitive support \cite{liu2026metaclass}, and hybrid human-AI regulation \cite{song2024exploreself,li2025flora,ge2025srlagent}. These systems demonstrate the potential of AI to support planning, monitoring, and reflection, but are typically studied within task- or course-bounded contexts.
Recent work suggests that chatbot-based systems can support multiple SRL phases \cite{guan2025chatbots,lee2025chatbotsrl,chatbotWTC2025}. However, they rarely maintain continuity across sessions or resources. This limitation becomes critical in informal digital learning, where learners must coordinate learning across distributed content and evolving goals \cite{informalprogramming2023, chatbotWTC2025}.

\subsection{Interactive Video-Based Learning Systems}

Video-based learning is central to informal education \cite{giannakos2013,sablic2021}. Platforms such as YouTube enable large-scale self-directed exploration \cite{lange2019,pires2022}, but their unstructured nature makes it difficult to maintain progression and conceptual coherence \cite{lee2017}. 
Prior work has explored concept-based navigation across video corpora. \textit{ConceptScape} enables collaborative concept mapping \cite{liu2018conceptscape}, and \textit{ConceptGuide} recommends structured navigation paths based on concept relationships \cite{tang2021conceptguide}. These systems highlight the importance of conceptual structure but focus on navigation within fixed corpora rather than learner-driven pathway construction. 
Recent LLM-driven systems focus on interaction within individual videos. \textit{Tutorly} supports apprenticeship-style learning \cite{li2024tutorly}, \textit{Untwist} enables multimodal question answering \cite{goudarzi2025untwist}, and \textit{Vid2Coach} transforms videos into task-oriented assistants \cite{vid2coach2025}. While these systems improve engagement and local comprehension, they primarily operate at the level of individual videos or sessions. 
Empirical work shows that video learning benefits from reflection and interaction \cite{navarrete2025,sablic2021}, but support for organizing multiple videos into coherent learning trajectories remains limited. Learners must still manually connect videos, track progression, and maintain context across sessions.

\subsection{AI Support for Planning and Learning Workflows}

Large language models have enabled new approaches to learning planning and workflow support. Planning-oriented systems such as \textit{PlanGlow} generate structured study plans \cite{chun2025}, while other work explores chaining LLMs for tutoring and instruction \cite{chen2023empowering}. Multi-agent approaches such as EduPlanner model learning trajectories using structured representations \cite{zhang2025eduplanner}. These systems primarily strengthen the forethought phase of SRL.
Complementary work focuses on learning-phase support. \textit{LearnMate} and \textit{CoGrader} provide contextual and evaluative assistance \cite{wang2025learnmate, cograder2025}, while \textit{Understood} supports real-time cognitive assistance \cite{understood2025}. Design frameworks further emphasize goal setting, feedback, and personalization \cite{chang2023designprinciples}. Reviews show that chatbot-based systems can support multiple SRL phases and improve confidence and performance \cite{guan2025chatbots,lee2025chatbotsrl}. 
However, these systems are typically studied within bounded tasks or structured settings and rarely maintain an explicit representation of learning progression across planning and learning. Even when conversational memory is available, it is often unstructured and not aligned with an explicit learning trajectory.

\subsection{Persistent Context and Cross-Session Learning Support}

Maintaining continuity across sessions and resources remains a central challenge in informal learning. Prior work highlights the importance of supporting SRL across distributed contexts. MetaTutor demonstrates the need to integrate multiple learning processes over time \cite{azevedo2022metatutor}, while MetaCLASS shows that LLMs struggle to sustain coherent pedagogical trajectories \cite{liu2026metaclass}. Studies of human-AI collaboration further emphasize shared representations of learner state and persistent context \cite{jarvela2023, zhang2026}. 
Despite these advances, most systems remain interaction-fragmented. Chatbots often lose context over time \cite{guan2025chatbots}, video systems focus on single-resource interactions \cite{li2024tutorly,goudarzi2025untwist}, and planning tools rarely carry structure into learning \cite{chun2025,wang2025learnmate}. As a result, learners must bridge planning and execution themselves, particularly in informal environments such as YouTube.

\textbf{Commercial AI learning systems.}
Even recent systems such as Google’s \textit{AboutLearn}~\cite{aboutlearn2024} and \textit{NotebookLM}~\cite{notebooklm2024} explore AI-assisted learning across multiple resources. AboutLearn generates topic-based video lists with per-video key terms and conversational support, while NotebookLM enables cross-resource organization and reasoning over user-provided materials. While these systems support multi-resource interaction, they remain largely \textit{collection-oriented} and lack a persistent interaction structure, organizing interactions around individual videos, documents, or chat threads. They do not maintain an explicit structure that represents progression or learner state across resources. As a result, users must reconstruct context and relationships over time.

\textbf{Summary and positioning.}
Prior work has advanced SRL support, video-based learning, and AI-driven planning, but these capabilities are often explored in isolation and lack a shared interaction structure. As a result, learners must bridge planning, progression, and learning execution themselves, particularly in YouTube-based informal learning. We argue that this limitation stems not only from lack of integration, but from the absence of an explicit structure for organizing interaction across resources in a way that supports self-regulated learning across phases. Existing systems treat plans, recommendations, or conversations as outputs, rather than persistent structures that shape interaction. We address this gap by conceptualizing the learning pathway as an \textit{explicit interaction structure} grounded in SRL that persists across phases and organizes how users interpret, navigate, and act on learning resources in YouTube-based learning. This reframes the design space from generating better outputs to structuring interaction itself.

\section{Formative Study}

To inform the design of YT-Pilot, we conducted a formative study to examine how learners currently use LLM-supported tools and YouTube's learning channel to construct informal learning plans. We focused on how these tools supported conceptual progression and longer-term learning organization. Five PhD students in computer science and information science from a mid-sized college town participated in the study (see Table~\ref{tab:formative} in the Appendix).

\subsection{Procedure and Analysis}

All sessions were conducted remotely (around one hour) and were approved by the institutional IRB. Participants first completed a 40-minute task in which they selected a topic of personal interest and constructed a YouTube-based learning plan using two conditions in counterbalanced order: (1) an LLM-supported environment (ChatGPT and Gemini) and (2) YouTube's existing learning channel. During each condition, participants were asked to think-aloud while reflecting on how well the generated plans aligned with their goals, preferences, and expectations, including comments on structure, sequencing, and the perceived usefulness of recommended videos.
The task was followed by a 20-minute semi-structured interview on informal learning practices, challenges, and unmet needs, with particular attention to goal setting, knowledge construction, progression across videos, and desired future features (see Appendix~\ref{app:interview-questions}).

Audio and video recordings were transcribed and analyzed using thematic analysis~\cite{braun2006}. Three researchers independently conducted initial coding, then iteratively refined the codebook and reconciled differences through inter-coder and intra-coder agreement checks. Final themes were consolidated through repeated rounds of collaborative review. We also collected and analyzed participants' prompts to better understand how LLM prompting, goal-setting, and planning can be structured to support learners' needs.

\subsection{Formative Findings}

Participants identified four recurring challenges in constructing informal YouTube learning plans with LLM-supported tools and YouTube's learning channel.

\textbf{Lack of structured sequencing and conceptual coherence.}
Participants consistently reported difficulty understanding how videos related to one another within a broader learning progression. Although LLMs could produce sequential plans, the conceptual relationships between recommended videos were often unclear. Rather than functioning as parts of a progressive pathway, videos often felt self-contained. As P2 explained, \textit{``each of these videos is a learning plan in itself\ldots{} it's not chunks of videos that together form a learning plan.''} Participants expressed similar concerns about YouTube's Learning channel. While they appreciated its high-level topical organization, they found the sequencing between categories difficult to interpret, with P2 noting, \textit{``I don't see how one leads to another.''} P4 likewise described the generated plans as overwhelming and poorly structured: \textit{``it's like it's giving you everything all at once, and it's not very well structured.''} These observations suggest existing tools do not clearly communicate how concepts build over time.

\textbf{Limited support for visualizing and navigating learning pathways.}
Participants also struggled to form an overall view of the learning space and navigate it effectively. Long, text-heavy plans were difficult to follow, and many participants wanted more visual representations, such as tables, graphs, or knowledge maps. They emphasized the need to understand the overall structure before diving into individual topics. For example, P1 described a preference to \textit{``start from the whole image, and then dive deep into different branches.''} Others found dense plans confusing, with P4 noting that \textit{``it just makes me confused\ldots{} giving me everything all at once.''} Participants explicitly requested more structured and visual formats. As P3 put it, \textit{``I think I would love for this to have\ldots{} a tabular version, or, like, figures, or more visual content.''}

\textbf{Fragmented note-taking and lack of integrated learning records.}
Participants reported that note-taking was poorly supported within existing tools. Some relied on external documents to track what they had learned. P2 noted, \textit{``I write it down, whether it's, like, a Google Doc,''} and P3 similarly explained, \textit{``I will track whatever thoughts that I have\ldots{} and write them there.''} Others did not maintain records consistently, making it harder to retain continuity across sessions and sometimes leading to relearning. As P5 described, \textit{``there's no point in me trying to track things I did if I forget it.''} These findings highlight the lack of integrated support for preserving learning context.

\textbf{Lack of persistent context in LLM-supported learning.}
Participants also expressed frustration with the absence of long-term memory in LLM-supported environments. Learning context was often lost across extended interactions, forcing them to restate prior information. This disrupted learning flow and added cognitive burden. P1 described this limitation: \textit{``if I spend a really long time with it\ldots{} at some point it cannot remember what we did previously\ldots{} and that would be frustrating.''} This finding points to the need for systems that preserve continuity across learning sessions.

\section{\systemName: Connected Workflow for Planning and Learning}

We instantiate our SRLT-grounded design through YT-Pilot, an interactive system that connects planning and learning through a persistent interactive pathway representation. The pathway is first constructed during planning and then carried forward into learning, where it becomes the basis for navigation, progress tracking, pathway-aware assistant and context-based note-taking.

\begin{figure*}
    \centering
    \includegraphics[width=\linewidth]{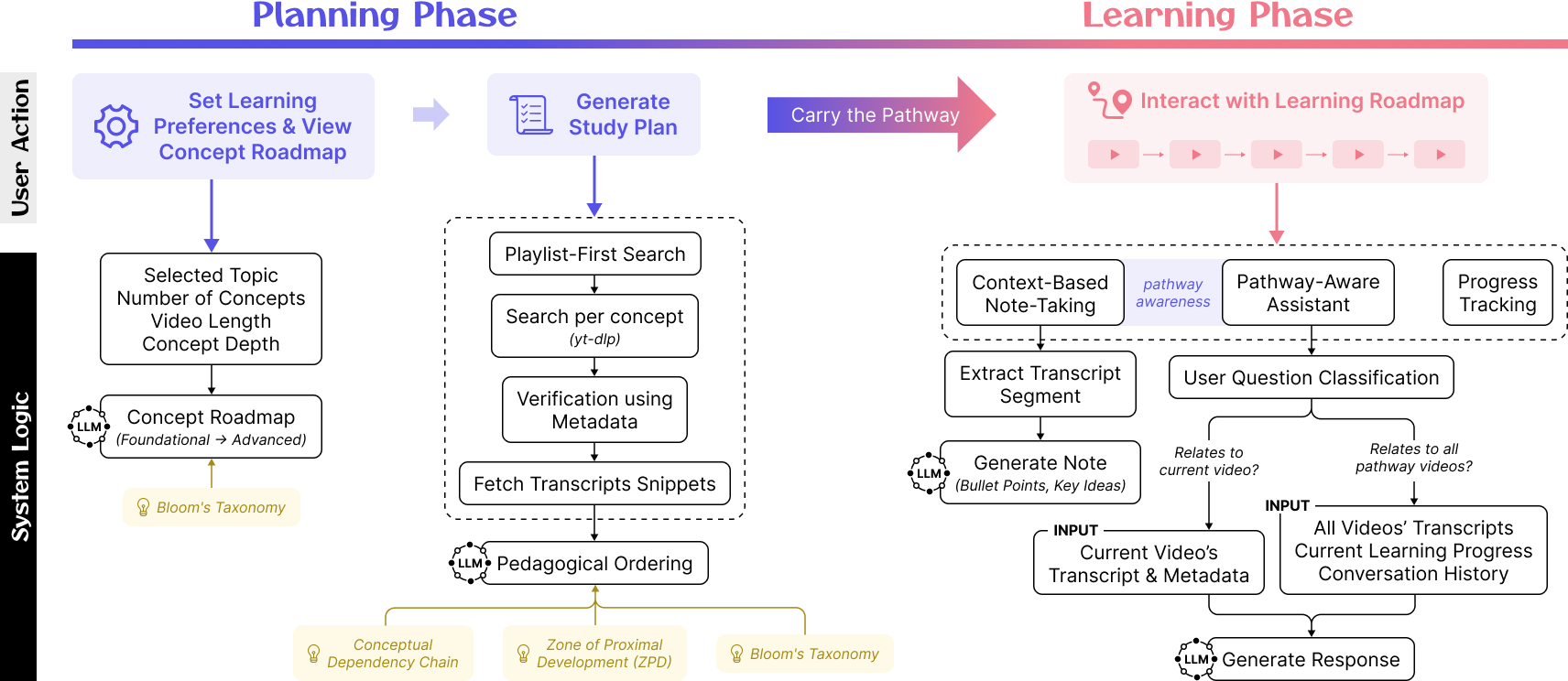}
    \caption{System architecture for YT-Pilot across planning (left) and learning (right) phases, with applied theories in yellow.}
    \Description{Two-phase system architecture diagram. Left side shows the planning phase: user inputs flow to concept roadmap generation via Bloom's Taxonomy, then to playlist-first search, per-concept yt-dlp search, metadata verification, transcript snippet fetching, and LLM pedagogical ordering using conceptual dependency chains, ZPD, and Bloom's Taxonomy. Right side shows the learning phase: user interacts with the learning roadmap; transcripts are fetched and cached; Context-based note-taking extracts transcript segments and generates notes; the pathway-aware assistant classifies questions and routes to current-video or pathway-wide context before generating responses.}
    \label{fig:logic}
\end{figure*}

\subsection{Design Goals}

We derive three design goals centered on treating the learning pathway as a shared structure connecting planning and learning.

\textbf{DG1. Pathway-Oriented Planning.}
Support goal setting and conceptual planning through structured preferences and concept preview, enabling learners to configure learning trajectories before committing to a pathway.

\textbf{DG2. Structured Multi-Video Progression.}
Generate pathways with explicit conceptual dependencies and inspectable progression, enabling learners to understand how videos build on each other and to revise the pathway as needed.

\textbf{DG3. Pathway-Aware Learning Support.}
Coordinate navigation, pathway-aware assistant and context-based note-taking through a persistent pathway representation, enabling learners to track progress and reason across videos during learning.

\subsection{Technical Architecture}

\systemName\ is a web-based system that supports pathway construction, navigation, and AI-assisted learning through a unified workflow. Central to the design is the learning pathway as a persistent computational object that connects planning and learning phases.

The system constructs pathways from learner preferences and concept structures, and uses this pathway to coordinate navigation, progress tracking, pathway-aware AI assistance, and context-based note-taking. The system is organized into two phases: a planning phase that constructs the learning pathway from user inputs and concept structures, and a learning phase that uses this pathway to coordinate interaction and learning activities.

Figure~\ref{fig:logic} provides an overview of the system architecture, illustrating how planning and learning are connected through the shared pathway representation. Detailed system logic for each phase is described in the following subsections.


\subsection{Learning Preferences and Concept Preview}

\begin{figure}[h]
  \centering
  \includegraphics[width=\columnwidth]{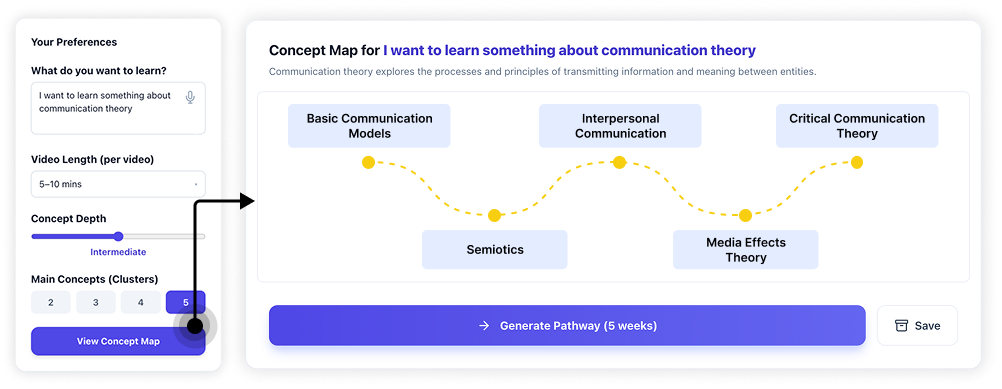}
  \caption{Concept preview enables learners to specify learning goals and preferences, then inspect an AI-generated concept map that provides a high-level, conceptual breakdown of the topic before pathway generation.}
  \Description{Screenshot of the concept preview stage in YT-Pilot. On the left is a preference panel where the learner enters a topic about communication theory, selects preferred video length, adjusts concept depth, and chooses the number of main concept clusters. On the right is a concept map titled for learning communication theory, showing labeled clusters such as Basic Communication Models, Interpersonal Communication, Critical Communication Theory, Semiotics, and Media Effects Theory connected by a dotted path with circular markers. A large button below the map offers to generate a five-week pathway, with a secondary save button beside it.}
  \label{fig:concept-preview}
\end{figure}

Learners begin by entering a topic and specifying planning preferences (Figure~\ref{fig:concept-preview} left, Figure~\ref{fig:logic} left), including preferred video length, experience level, and the number of concept clusters to include in the pathway. \systemName\ then generates a concept map to help learners understand the topic before pathway generation. The map breaks down and visualizes the topic as a progression from foundational to more advanced concepts (Figure~\ref{fig:concept-preview} right, Figure~\ref{fig:logic} left). Generated through a structured LLM prompt informed by Bloom’s Taxonomy~\cite{bloom1956}, it organizes concepts into a learnable sequence (see Appendix~\ref{app:prompt-concept-map}). This map serves as the structural backbone for pathway generation, where each concept cluster is instantiated into one or more videos, providing a concrete basis for shaping the study plan.

\subsection{Learning Pathway Generation}

After learners confirm their planning preferences, \systemName\ generates a structured study plan that organizes videos into a coherent learning pathway (Figure~\ref{fig:logic} middle). The pathway is constructed by mapping videos to concepts in the roadmap and organizing them into a dependency-aware sequence, where each video contributes to pathway progression. The generation process is informed by conceptual dependency chains~\cite{gagne1985}, the Zone of Proximal Development~\cite{vygotsky1978}, and Bloom’s Taxonomy~\cite{bloom1956}. The resulting pathway is organized by weeks and includes learning objectives, conceptual dependencies, video-level rationale, and core metadata. Rather than simply recommending relevant videos, the system structures them into a progression that learners can inspect, interpret, and follow over time. Details on retrieval, filtering, and pedagogical ordering are provided in Appendix~\ref{app:system-prompts} and illustrated in Figure~\ref{fig:pathway_logic}.

\subsection{Study Plan Review and Course Enrollment}

\begin{figure}
  \centering
  \includegraphics[width=\columnwidth]{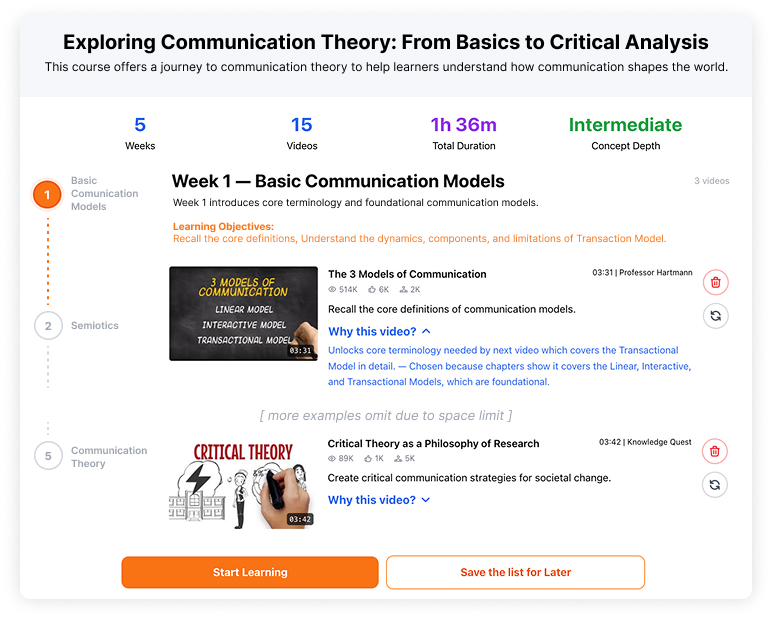}
  \caption{Pathway review presents a structured multi-week learning plan with ordered videos, learning objectives, and per-video rationales to help learners understand and revise the generated pathway before starting.}
  \Description{Screenshot of the pathway review stage in YT-Pilot for a course on communication theory. The interface shows a weekly learning plan titled Week 1, Basic Communication Models, with learning objectives listed beneath the heading. A vertical sequence on the left marks the progression of concept clusters across the pathway. The main panel displays individual video entries with thumbnail, title, metadata, brief instructional purpose, and an expandable explanation of why the video was selected. Action buttons at the bottom allow the learner to start learning or save the list for later.}
  \label{fig:pathway-review}
\end{figure}

Once the pathway is generated, learners review it before starting the learning phase (Figure~\ref{fig:pathway-review}). The interface presents the overall structure, including course description, duration, and a weekly breakdown with learning objectives and ordered videos. For each video, learners can inspect a ``Why this video?'' explanation that clarifies its role within the pathway. They can also modify the pathway by removing or replacing individual videos while preserving the overall structure. This step allows learners to inspect, interpret, and refine the pathway before committing to it, supporting alignment between their goals and the generated learning trajectory. Once satisfied, they transition into the learning environment.

\subsection{Learning Space}

\begin{figure*}[t]
  \centering
  \includegraphics[width=\linewidth]{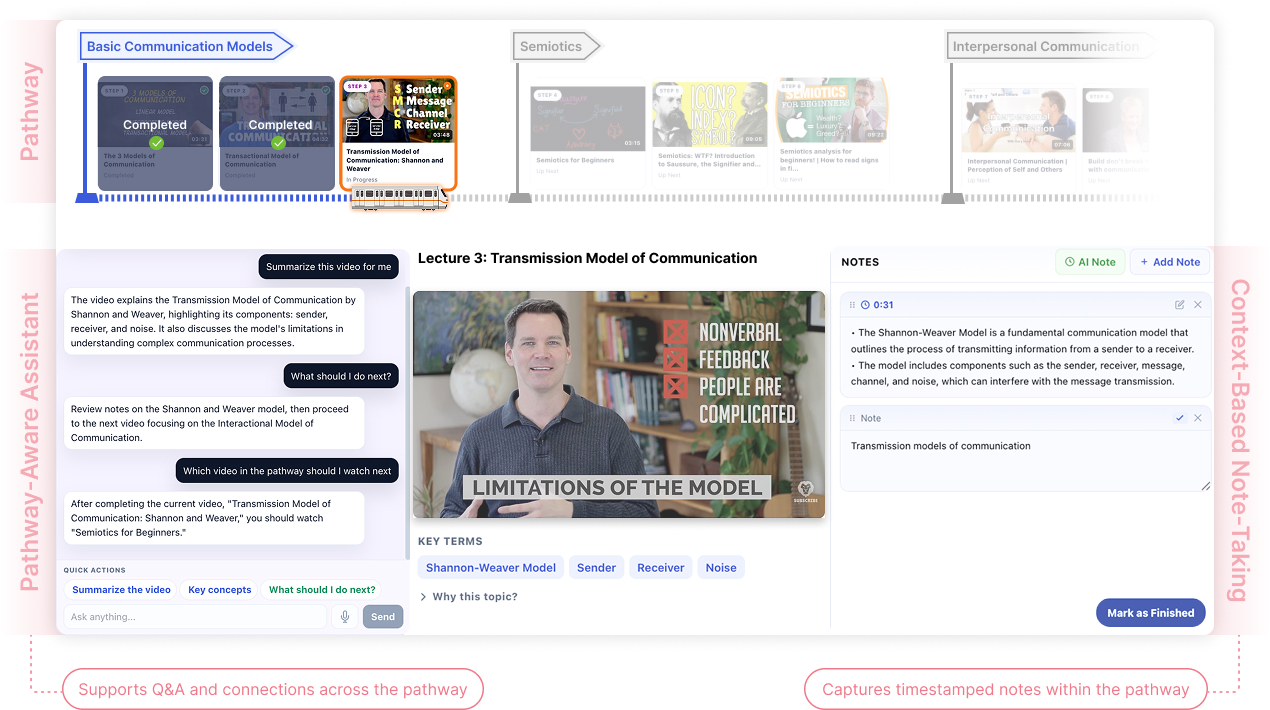}
  \caption{Learning environment for video study includes: (a) concept map and generated learning pathway (top), (b) main learning space that integrates the instructional video, pathway-aware conversational guidance, key terms, and a context-based note-taking panel to support focused learning within a structured pathway (bottom).}
  \Description{Composite interface view showing two connected stages of YT-Pilot. The upper panel presents a concept map for communication theory, with labeled clusters including Basic Communication Models, Interpersonal Communication, Critical Communication Theory, Semiotics, and Media Effects Theory, connected by a dotted progression line. The lower panel shows the resulting pathway as a horizontal timeline of video cards grouped by concept cluster, with the currently selected video highlighted and later topics faded in the distance.}
  \label{fig:learning}
\end{figure*}

In the learning phase, the pathway becomes the primary interaction structure (Figure~\ref{fig:learning} top). It is presented as an interactive roadmap in which conceptual clusters serve as milestone stations connected by a rail, with a moving train indicating the learner’s current progress. Corresponding videos are sequenced along the pathway to support progress tracking and navigation. The main learning space integrates the video player, pathway-aware assistant, and context-based note-taking panel (Figure~\ref{fig:learning} bottom), supporting learners' understanding, efficiency, and engagement with the content, as detailed below.

\subsection{Pathway-Aware Assistant}
\systemName\ includes a pathway-aware assistant that operates over the learning pathway, using it as a structured context to support both local video understanding and cross-video reasoning (Figure~\ref{fig:learning}, lower left, Figure~\ref{fig:logic} right). When a pathway is initialized, the system prepares transcripts and metadata context across all videos, enabling the assistant to draw on the broader learning trajectory rather than a single video. To support different forms of help, the assistant uses a pathway-grounded context assembly mechanism that incorporates the learner’s current position, prior progress, and conceptual relationships across videos. It also classifies incoming questions based on whether they concern the current video or the broader pathway. Current-video questions are answered using the active video's transcript and metadata, while pathway-level questions draw on aggregated context from multiple videos, pathway structure, progress, and learner history. This design enables the assistant to support both local clarification and broader synthesis across the learning trajectory. Further details are provided in Appendix~\ref{app:system-prompts} and illustrated in Figure~\ref{fig:aiassistant_logic}.

\subsection{Context-Based Note-Taking}
The note-taking panel supports both free-form notes and AI-generated notes tied to the current video timestamp (Figure~\ref{fig:learning} lower right, Figure~\ref{fig:logic} right). Integrated into the pathway-based workflow, it supports the shift from watching to reflection within SRLT. When triggered, the system generates concise timestamped notes from the transcript around the learner's current playback position. AI notes function as lightweight scaffolds for capturing key information, while manual notes allow learners to record their own interpretations and actively engage with the content. As notes are anchored to specific positions within the pathway, they help preserve context and maintain continuity across videos without requiring manual transfer. By remaining embedded within the pathway, notes contribute to continuity across the broader learning process. Yet, this design introduces a trade-off between efficiency and engagement: AI-generated notes reduce effort, but may limit deeper cognitive processing, which is why the system supports manual note-taking as a complementary mechanism for reflection and deeper engagement. Further details on note generation and anti-repetition design are provided in Appendix~\ref{app:system-prompts} and illustrated in Figure~\ref{fig:notetaking_logic}.

\section{User Evaluation}
To examine how \systemName’s pathway-based interaction shaped learners' perceptions relative to YouTube's existing learning channel, we conducted a within-subjects study comparing the two systems across planning, pathway generation, and learning support.

\subsection{Study Design and Participants}
\label{sec:participants}

We used a within-subjects design to enable direct comparison between \systemName\ and YouTube Learning while controlling for individual differences in prior knowledge and learning habits. Participants interacted with both systems in counterbalanced order to mitigate order effects. We selected YouTube Learning as the baseline because it represents YouTube’s own initiative for structured informal learning, providing topic-organized video lists based on user-specified topics and an AI assistant. This choice increases ecological validity by grounding the comparison in a platform that participants already use for informal learning, rather than an artificial or unfamiliar baseline~\cite{shadish2002experimental}.

We initially recruited 23 participants through a U.S. university mailing list and Slack workspace. We excluded 3 from analysis, one due to inattentive task responses and two due to system downtime, resulting in a final sample of 20 (11 male, 9 female). All participants aged 18--34 years. 19 participants were students across different degrees, and 1 participant was an early-career professional. Fields of study included computer science, information management, HCI, informatics, and computer engineering (see Table~\ref{tab:participant-demographics} in Appendix). All participants reported prior experience with YouTube-based informal learning ($M = 4.50$ on a 7-point scale) and high familiarity with AI tools ($M = 6.45$, $SD = 0.83$). The study was approved by the university IRB, and participants received \$25 compensation.

\subsection{Study Procedure}
\label{sec:procedure}
Each study session lasted approximately 70 minutes and was conducted remotely via Zoom, with a researcher present to observe interactions and provide navigation guidance when needed. The procedure consisted of four phases.

\textbf{Phase 1 --- Consent, overview, and pre-survey (10 min):}
Participants first reviewed the study objectives, procedures, and data-handling protocol and provided informed consent online. Then, they received a brief overview of the two systems and completed a short pre-survey covering demographics, YouTube learning experience, and familiarity with AI-assisted tools (see Appendix~\ref{app:pre-survey}).

\textbf{Phase 2 --- System interaction (40 min; 20 min per condition):}
Participants were assigned to one of two counterbalanced orders, \systemName\ first or YouTube Learning first. Before each condition, the researcher briefly introduced the key features of the system. Participants then selected a topic they were personally interested in learning, and used the same topic in both conditions. In the \systemName\ condition, participants set learning preferences, previewed the concept map, generated a learning pathway, reviewed and modified the plan, and then used the learning environment, including the video roadmap, pathway-aware assistant, and context-based note-taking features. They were encouraged to explore the pathway, watch portions of videos, ask questions to the assistant, and generate notes. In the YouTube Learning condition, participants searched for the same topic, explored the generated learning journey, browsed the organized video lists, interacted with YouTube's built-in AI assistant, and watched portions of recommended videos.

\textbf{Phase 3 --- Post-survey (10 min):}
After completing both conditions, participants evaluated each system's support for planning, pathway structure, and learning experience using 12 Likert-scale items. The items were adapted from established instruments, including the Technology Acceptance Model~\cite{davis1989}, with slight modifications to align with our system’s goals, as well as goal-setting scales~\cite{hew2025} and the explainability and controllability framework~\cite{chun2025} (see Appendix~\ref{app:post-survey}).

\textbf{Phase 4 --- Semi-structured interview (10 min):}
Participants reflected on their learning processes, features they valued or found lacking, and unmet needs in guided discussion in two systems.

\subsection{Measures and Analysis}
\label{sec:data-collection}

We collected quantitative data through post-surveys and qualitative data through semi-structured interviews. For quantitative analysis, we used Wilcoxon signed-rank tests to compare within-subject ratings between \systemName\ and YouTube Learning on shared survey items. Items unique to \systemName\ are reported separately. For qualitative analysis, interview recordings were transcribed and analyzed using thematic analysis~\cite{braun2006}. Three researchers iteratively coded the transcripts, identified recurring themes, and organized the findings around the research questions.

\section{Results}

We report post-survey, interaction log, and interview results, focusing on how \systemName’s pathway-based design shaped learners’ ability to construct, interpret, and navigate learning trajectories across planning, pathway generation, and learning phases, compared to YouTube Learning.

\subsection{Behavioral Overview of Learning-Phase Interactions}
To contextualize the survey and interview findings, we analyzed interaction logs from the \systemName\ condition, focusing on use of the assistant and note-taking features.

\textbf{Pathway-aware assistant engagement.}
All participants used the pathway-aware assistant. On average, they sent $M = 15.2$ messages ($SD = 12.7$, range: 2--52) and received $M = 21.5$ responses ($SD = 16.0$). Analysis of 147 unique user messages identified 6 interaction categories. Quick-action buttons (\textit{Summarize}, \textit{Key Concepts}, \textit{What Should I Do Next}) accounted for 25\% of queries, suggesting that low-effort entry points supported assistant engagement. Notably, 17\% of queries were pathway-level, referring to other videos, asking about cross-video relationships, or requesting navigation across the learning trajectory. Examples included \textit{``which video explains how decision trees deal with regression?''} (P10), \textit{``how does this video connect to prior videos in the study plan?''} (P9), and \textit{``can you read the videos in my learning pathway and explain the relation between each video?''} (P5). An additional 14\% of queries focused on content within the current video. These patterns suggest that learners used the assistant not only for local comprehension, but also to reason across videos and navigate the pathway as a structured learning trajectory.

\textbf{Note-taking patterns.}
All participants generated at least one AI note ($M = 4.5$, $SD = 3.0$), and 15 participants (75\%) also added manual notes alongside AI-generated ones ($M = 1.4$, $SD = 2.2$). In total, 89 AI-generated notes and 27 manual notes were created. AI notes were timestamp-anchored summaries tied to specific videos within the pathway, while manual notes ranged from short factual reminders (e.g., \textit{``SaaS: Software as a service''}) to broader conceptual observations. The combination of AI and manual notes suggests that note-taking functioned as a lightweight scaffold for capturing content within the pathway context, while still allowing learners to construct their own interpretations. By anchoring notes to specific positions in the pathway, this design supported continuity across videos without requiring learners to manually transfer information between learning steps.

\subsection{RQ1: Goal Setting and Conceptual Connectivity in the Planning Phase}

Table~\ref{tab:rq1} presents participants' evaluations of the planning phase. Overall, results favored \systemName\ across all measures. Compared with YouTube Learning, \systemName\ received significantly higher ratings for helping participants define their learning goals clearly and reflect on a personal learning plan. The YT-Pilot-only concept roadmap item also received a high rating, indicating its usefulness for understanding the topic structure before pathway generation.

\begin{table}[h]
\centering
\caption{Perceived Quality of Planning-Phase Support}
\label{tab:rq1}
\small
\begin{tabular}{p{3.5cm}p{1.2cm}p{1.2cm}p{0.6cm}p{0.2cm}}
\toprule
\textbf{Item (rating 1-5)} & \systemName\ & \textbf{YouTube} & $W$ & $r$ \\
\midrule
Define learning goals clearly
  & \textbf{4.45} [0.76] & 3.25 [1.21] & 16** & .79 \\
Reflect on personal plan
  & \textbf{4.35} [0.81] & 2.65 [1.09] & 4*** & .95 \\
\midrule
\multicolumn{5}{l}{\textit{YT-Pilot only:}} \\
Concept roadmap (big picture)
  & \multicolumn{2}{l}{4.55 [0.76]} & \multicolumn{2}{c}{---} \\
\bottomrule
\multicolumn{5}{l}{\scriptsize *$p < .05$, **$p < .01$, ***$p < .001$.Mean [SD]. $r$ = rank-biserial correlation.}
\end{tabular}
\end{table}

\textbf{Structured planning controls supported pathway configuration (T1).}
Participants rated \systemName\ significantly higher than YouTube Learning on defining learning goals clearly ($W = 16$, $p = .004$, $r = .79$), with higher ratings for \systemName\ ($M = 4.45$, $SD = 0.76$ vs.\ $M = 3.25$, $SD = 1.21$). Interview data suggest that this was driven by the structured preference form, which enabled learners to configure key aspects of the learning pathway, such as video length, experience level, and concept coverage. Participants linked this to greater control over how the pathway was constructed. P1 noted, \textit{``I think there's more customisations and adjustments I can make\ldots{} I can actually adjust the video time\ldots{} less control''} in YouTube Learning, while P6 emphasized that \textit{``the YouTube Learning does not allow me to define my own goals at all.''}. These findings suggest that planning support extends beyond goal definition to shaping the structure of the resulting pathway.

\textbf{Concept roadmap supported early orientation (T2).}
Participants responded positively to the concept roadmap ($M = 4.55$, $SD = 0.76$), which provided a big-picture view before pathway generation. Interview data suggest this supported early understanding of scope and structure. P1 noted, \textit{``I really need to know the big picture before I start,''} while P4 described it as helping them \textit{``see\ldots{} the big picture.''}. Together, these responses suggest that the roadmap supported conceptual orientation early in the planning process, making it easier for learners to understand the scope of the topic and approach pathway generation with greater confidence.

\subsection{RQ2: Structured Learning Progression in Pathway Generation}
Table~\ref{tab:rq2} summarizes participants' evaluations of pathway generation. Overall, results indicate that \systemName\ produced a more coherent and inspectable learning pathway than YouTube Learning. Compared with YouTube Learning, \systemName\ was rated higher on pathway clarity, logical connection across videos, understanding why videos were included, and confidence in revising the pathway, while ratings for meeting goals and needs did not differ significantly. The YT-Pilot-only item on weekly task connections also received a strong rating, suggesting that the week-level organization supported perceptions of a connected study plan.

\begin{table}[h]
\centering
\caption{Perceived Quality of Pathway Generation}
\label{tab:rq2}
\small
\begin{tabular}{p{3.4cm}p{1.2cm}p{1.2cm}c c}
\toprule
\textbf{Item (rating 1-7)} & \systemName\ & \textbf{YouTube} & $W$ & $r$ \\
\midrule
Pathway clearly presented
  & \textbf{6.45} [0.76] & 5.10 [1.29] & 8*** & .90 \\
Pathway met goals \& needs
  & \textbf{5.55} [1.32] & 4.85 [1.39] & 42 & .45 \\
Videos logically connected
  & \textbf{5.75} [1.21] & 4.40 [1.43] & 21** & .73 \\
Confident to revise pathway
  & \textbf{5.15} [1.42] & 3.60 [1.82] & 18* & .73 \\
Understand why included
  & \textbf{5.90} [0.97] & 3.85 [1.63] & 9*** & .89 \\
\midrule
\multicolumn{5}{l}{\textit{YT-Pilot only:}} \\
Weekly task connections clear
  & \multicolumn{2}{l}{5.90 [1.37]} & \multicolumn{2}{c}{---} \\
\bottomrule
\end{tabular}
\end{table}

\textbf{Structured progression improved perceived coherence over cluster-based organization (T3).}
Participants perceived \systemName\ as more coherent than YouTube Learning ($W = 21$, $p = .008$, $r = .73$), with higher ratings for logical connection across videos ($M = 5.75$ vs.\ $M = 4.40$). Interview data suggest this difference stemmed from presenting the pathway as a progressive sequence rather than topic clusters. P1 described \systemName\ as \textit{``more regulated''} and compared it to a curated series that moves through \textit{``block A, B, C, D, E on one topic.''} P4 similarly noted \textit{``I could see that, okay, this is in progression over time,''} whereas in YouTube Learning \textit{``it was harder to understand what the progression was supposed to be.''}. In contrast, participants described YouTube Learning as grouping related content without clearly showing how one video led to the next. P15 referred to \textit{``more clusters''} than \textit{``logical lines,''} and P17 similarly described it as \textit{``clusters rather than progression.''} These findings suggest that \systemName\ makes progression visible at the pathway level.

\textbf{Targeted pathway editing increased confidence, though revision controls remained limited (T4).}
Participants reported greater confidence in revising pathways with \systemName\ ($W = 18$, $p = .010$, $r = .73$), with higher ratings ($M = 5.15$ vs.\ $M = 3.60$). This reflects the value of making targeted changes while preserving structure. P4 noted \textit{``I could regenerate videos,''} while P17 emphasized \textit{``it's a targeted change at one video.''} However, limitations remained; P5, for example, requested controls to reorder videos. These findings suggest that pathway effectiveness depends on both initial structure and the ability to revise it in a focused and understandable way.

\textbf{Rationale-driven explainability strengthened understanding of the pathway structure (T5).}
Participants also rated \systemName\ significantly higher than YouTube Learning on understanding why videos were included ($M = 5.90$ vs.\ $M = 3.85$; $W = 9$, $p = .001$). Qualitative data suggest that this difference was driven by the \textit{Why This Video} feature, which helped participants understand both why each video was selected and how it fit into the broader pathway. P1 noted that \systemName\ did a better job \textit{``explaining why each video was there, and linking the layers between different videos.''} P4 similarly said that these explanations were helpful because \textit{``I could understand why it added it.''} P12 further described the rationale as useful when deciding whether to modify the plan, explaining that it \textit{``gave me more confidence to replace the video with something else.''} These findings suggest that explainability was an important mechanism through which \systemName\ made the generated pathway feel more coherent and trustworthy.

\subsection{RQ3: Progress Monitoring and Contextual Support in the Learning Phase}

Table~\ref{tab:rq3} presents participants' evaluations of the learning phase, focusing on how \systemName\ supported navigation and progression along the learning pathway. Overall, \systemName\ was rated more positively for supporting the broader learning process, particularly in managing the pathway, tracking progress, and supporting learning. In contrast, YouTube Learning was rated higher for AI assistant helpfulness. These results suggest a distinction between pathway-level support and video-level assistance, with \systemName\ emphasizing structured progression and cross-video coordination, and YouTube Learning providing more localized support within individual videos.

\begin{table}[h]
\centering
\caption{Perceived Quality of Learning-Phase Support}
\label{tab:rq3}
\small
\begin{tabular}{p{3.8cm}p{1.2cm}p{1.2cm}p{0.3cm}p{0.3cm}}
\toprule
\textbf{Item (rating 1-5)} & \systemName\ & \textbf{YouTube} & $W$ & $r$ \\
\midrule
Improved ability to learn
  & \textbf{4.45} [0.60] & 3.75 [1.07] & 16* & .69 \\
AI roadmap easier to manage
  & \textbf{4.50} [0.69] & 3.50 [0.95] & 0*** & 1.00 \\
AI assistant helpful
  & 3.20 [0.89] & \textbf{4.30} [0.80] & 10** & .87 \\
Progress tracking
  & \textbf{4.30} [0.86] & 2.95 [1.19] & 4** & .93 \\
System easy to use
  & \textbf{4.35} [0.81] & 3.90 [1.07] & 6 & .67 \\
\midrule
\multicolumn{5}{l}{\textit{YT-Pilot only:}} \\
Context-based note-taking useful
  & \multicolumn{2}{l}{3.50 [1.40]} & \multicolumn{2}{c}{---} \\
\bottomrule
\end{tabular}
\end{table}

\textbf{Structured roadmap support improved perceived learning effectiveness (T6).}
\systemName\ received higher ratings than YouTube Learning on improved ability to learn ($M = 4.45$ vs.\ $M = 3.75$; $W = 16$, $p = .019$). Participants linked this difference to the way \systemName\ organized learning as a sequenced pathway rather than a set of loosely related videos. P12 described the system as feeling more grounded in how learning should be structured, noting that it was \textit{``systematising, based on, like, how learning happens''} and \textit{``suited towards more of an educational framework.''} This perception also appeared in how participants described moving through the pathway. For P11, the value of the system was that \textit{``it does show me that I've watched this video, and I need to go forward and learn the next concept.''} P19 similarly emphasized that without this kind of structure, it would be harder to know \textit{``I should learn this first, and that later.''} Taken together, these responses suggest that \systemName\ improved perceived learning support by making progression explicit and instructionally meaningful.

\textbf{Context-based note-taking supported lightweight capture but revealed a tension with active learning (T7).}
Responses to the note-taking feature were mixed, with a moderate rating ($M = 3.50$, $SD = 1.40$). Participants appreciated it as a lightweight way to capture content during videos. P1 valued that notes were \textit{``marked by timestamp''} and \textit{``listed by bullet points,''} describing them as \textit{``structured and easy to read.''}. At the same time, limitations highlighted gaps in depth and continuity. P9 noted that notes remained tied to the current video and did not carry across the pathway (\textit{``the notes don't show the previous notes''}), while P18 felt they \textit{``didn't go into, like, specifics.''} More critically, P17 emphasized the importance of active engagement, noting that \textit{``in the process of note-taking, you learn what's important.''}. These findings suggest that context-based note-taking functions as a lightweight scaffold within the learning phase, but does not fully support deeper reflection or cross-video continuity. This reveals a tension between efficient capture and active learning, highlighting the need to better integrate note-taking across the pathway to support reflection within the SRLT workflow.

\textbf{The pathway-aware assistant showed complementary strengths in pathway guidance and video-level precision (T8).}
In contrast to other learning-phase findings, the pathway-aware assistant was rated higher in YouTube Learning than in \systemName\ ($M = 4.30$ vs.\ $M = 3.20$; $W = 10$, $p = .001$). Interview data suggest that this reflects a difference in assistant scope rather than overall quality. Participants valued \systemName’s assistant for its pathway-level awareness, as P2 noted, \textit{``it had, like, previous context about other videos that I watched as well. It knew where I am, it knew what I was already learning.''} P10 similarly described using it to connect videos and decide what to watch next, stating, \textit{``the AI assistant can… make connections between different videos… so I can just ask, like, I want to learn this, and which video should I watch?''} In contrast, YouTube Learning’s assistant was perceived as stronger for precise, in-video support. As P6 explained, \textit{``[YouTube Learning AI] can directly generate a link… it has features where it redirected you to the specific timing or the time step in the video.''} Overall, these findings reveal a trade-off between persistent pathway-level guidance across videos and fine-grained, video-level precision within individual videos.

\textbf{Progress tracking improved orientation and motivation during learning (T9).}
A clear advantage of \systemName\ appeared in progress tracking, where it outperformed YouTube Learning ($M = 4.30$ vs.\ $M = 2.95$; $W = 4$, $p = .001$). Participants attributed this difference to the visual roadmap, completion state, and sense of forward movement built into the pathway. P1 described it as \textit{``It's, like, linking learning or Coursera kind of style that is really helpful for track the progress.''} P4 similarly emphasized its practical value, noting \textit{``it was just much easier to see, like, the sequence of videos. And I could… just, like, mark things as done.''} Others highlighted its motivational role; P7 explained that \textit{``And another thing is the YT pilots, they have the mark at the finish, right?… Give a sense of accomplishment.''} These findings suggest that progress tracking supported both orientation within the pathway and motivation during learning.

\textbf{Integrated support improved convenience, but introduced interface complexity (T10).}
Ease-of-use ratings did not differ significantly between conditions ($W = 6$, $p = .109$, $r = .67$), though \systemName\ was rated slightly higher ($M = 4.35$ vs.\ $M = 3.90$). Qualitative data explain this pattern. Participants valued having multiple forms of support in one interface; P11 noted, \textit{``as a user, I don't really need to change tabs… one good feature about YTPilot is that it's in-house. It's like integrating a lot of stuff.''} At the same time, this feature richness introduced complexity. P14 described that \textit{``there's so many different panels, and I think it's hard to keep track of.''} and P18 warned that \textit{``having too much, like, custom ability will make it harder for the users.''} Overall, the integrated design improved convenience but introduced a usability tension around layout complexity and cognitive load.

\section{Discussion}

We interpret our findings to understand how pathway-based interaction shapes informal video-based learning. We focus on how learners used the pathway across planning and learning phases, how progression was perceived, and how different forms of support were coordinated. We then relate these findings to prior work to clarify how persistent interaction structures can support continuity in informal learning.

\subsection{Interpreting Key Findings \& Design Implications}

\textbf{(1) The pathway as a persistent structure across planning and learning.}
Our findings show that learners relied on the pathway as a central reference across both planning and learning phases. During planning, structured preference input and concept roadmaps supported goal articulation and pathway configuration (RQ1). During learning, the same structure enabled navigation, progress tracking, and revisiting prior content (RQ3). Together, these results indicate that maintaining a persistent representation reduced the need to reconstruct context across videos, a common challenge in informal learning environments.

Prior work in self-regulated learning (SRL) emphasizes coordination between planning, monitoring, and reflection \cite{zimmerman2002}, while existing systems often support these phases in isolation or within bounded tasks \cite{chun2025,li2024tutorly,ge2025srlagent}. Our findings extend this line of work by showing that continuity in informal learning can be supported through a shared, persistent interaction structure that connects planning decisions with learning actions across distributed resources. Rather than treating plans as static outputs, the pathway functions as an ongoing structure that supports learners in maintaining context over time.

\textbf{Design Implication: Ground learning support in persistent structures.}
Our findings suggest that learning support should be grounded in persistent, user-facing structures that remain available across planning and learning. Maintaining such structures allows learners to interpret progress, revisit prior content, and coordinate learning actions over time.

\vspace{0.5em}

\textbf{(2) Making learning progression explicit in informal video-based learning.}
Participants consistently reported that \systemName\ made progression across videos more interpretable compared to cluster-based organization in YouTube Learning. Quantitative results (RQ2) showed higher ratings for pathway clarity, logical connection, and understanding why videos were included, while interview data indicated that learners perceived the pathway as a directional sequence rather than a collection of loosely related items.

Prior work has explored concept-based navigation and structured representations in video learning, such as ConceptScape and ConceptGuide \cite{liu2018conceptscape,tang2021conceptguide}. These systems highlight the importance of conceptual relationships but primarily support navigation within fixed corpora. Our findings extend this work by demonstrating that making progression explicit at the level of a learner-specific pathway—through dependencies, sequencing, and rationale—helps learners understand how knowledge builds over time. This suggests that supporting informal learning is not only about retrieving relevant content, but also about making progression visible, interpretable, and actionable.

\textbf{Design Implication: Make progression explicit and inspectable.}
Representing learning as a sequence of interdependent steps, rather than loosely related collections, supports learners in understanding how knowledge builds over time. Making dependencies, sequencing, and rationale visible enables more coherent navigation and interpretation.

\vspace{0.5em}

\textbf{(3) Coordinating pathway-level and local learning support.}
Our results reveal a distinction between pathway-level and video-level support in how learners engaged with AI assistance. Interaction logs showed that participants used the assistant for both localized clarification within a video and broader questions spanning multiple videos. While YouTube Learning was rated higher for immediate video-level assistance (RQ3), \systemName\ enabled cross-video reasoning and navigation by grounding assistance in the pathway structure.

Prior systems such as Tutorly focus on in-video assistance \cite{li2024tutorly}, while others such as SRLAgent and LearnMate support broader learning processes \cite{ge2025srlagent,wang2025learnmate}. Our findings extend this distinction by showing that these forms of support operate at different levels of scope and should be coordinated rather than treated as alternatives. By grounding interaction in a shared pathway representation, \systemName\ demonstrates how systems can support both local understanding and cross-video reasoning within a unified learning experience.

\textbf{Design Implication: Support interaction across multiple levels of scope.}
Learning involves both localized understanding within individual resources and broader reasoning across a trajectory. Systems should support both levels and coordinate them through shared context, enabling transitions between detailed exploration and higher-level navigation.

\vspace{0.5em}

\textbf{(4) Learning pathways as revisable rather than fixed structures.}
Participants reported higher confidence in revising pathways with \systemName\ compared to YouTube Learning (RQ2), particularly through targeted edits such as replacing individual videos. At the same time, they expressed the need for more flexible controls, such as reordering content or modifying structure more broadly. These findings suggest that learners actively interpret and adapt pathways rather than following them as fixed plans.

Prior work on learning planning systems has primarily focused on generating structured outputs \cite{chun2025,zhang2025eduplanner}, often treating plans as fixed recommendations. Our findings extend this perspective by showing that the effectiveness of structured learning depends not only on initial generation, but also on supporting ongoing inspection and revision. Designing pathways as revisable structures allows learners to maintain agency while benefiting from system-generated organization.

\textbf{Design Implication: Design learning structures as revisable.}
Learning pathways should function as evolving representations rather than fixed outputs. Supporting inspection and modification allows learners to adapt their trajectories as their understanding develops while maintaining structural guidance.

\subsection{Limitations and Future Work}
This study has several limitations. First, our sample consisted primarily of graduate students in technical fields at a single U.S.\ university, which limits the generalizability of the findings beyond this population, although this sampling strategy is consistent with prior work~\cite{chun2025,wang2025learnmate}. Second, each condition lasted approximately 20 minutes. This duration was sufficient for feature exploration, but it did not allow participants to complete an entire pathway or interact with the system across multiple sessions. Accordingly, our findings primarily reflect how pathway-based interaction supports perceived progression, navigation, and coherence within a single session, rather than long-term learning continuity across days or weeks. The value of progress tracking and pathway-level context may therefore differ under sustained use. Third, \systemName\ emphasizes a structured and goal-oriented workflow, which may not align with all forms of informal learning. Some learners may prefer lighter-weight or more exploratory interactions without explicit planning. Fourth, our evaluation compares \systemName\ primarily against YouTube’s Learning channel. While this provides a realistic and ecologically valid baseline, it does not capture the full range of emerging AI-supported learning tools. Additional comparisons with other systems, such as notebook-style or document-grounded assistants, may further contextualize the strengths and limitations of pathway-based interaction. Future work should therefore examine how pathway structure and interface support can adapt to different learner preferences and longer-term learning contexts.

\section{Conclusion}

We presented \systemName\, an AI-supported system that reframes informal YouTube learning as a connected, multi-phase workflow grounded in self-regulated learning theory. By centering interaction around a persistent learning pathway, the system links planning and learning while maintaining context across videos and interactions. In a within-subjects study ($N=20$) comparing \systemName\ with YouTube’s Learning channel, we found improvements in goal clarity, planning support, pathway coherence, and progress tracking, alongside a trade-off between pathway-level guidance and fine-grained video-level assistance. These findings highlight the value of pathway-centered interaction design for supporting more structured and self-regulated learning in informal video environments.

\appendix

\section{YouTube Learning}
\begin{figure*}
  \centering
  \includegraphics[width=\textwidth]{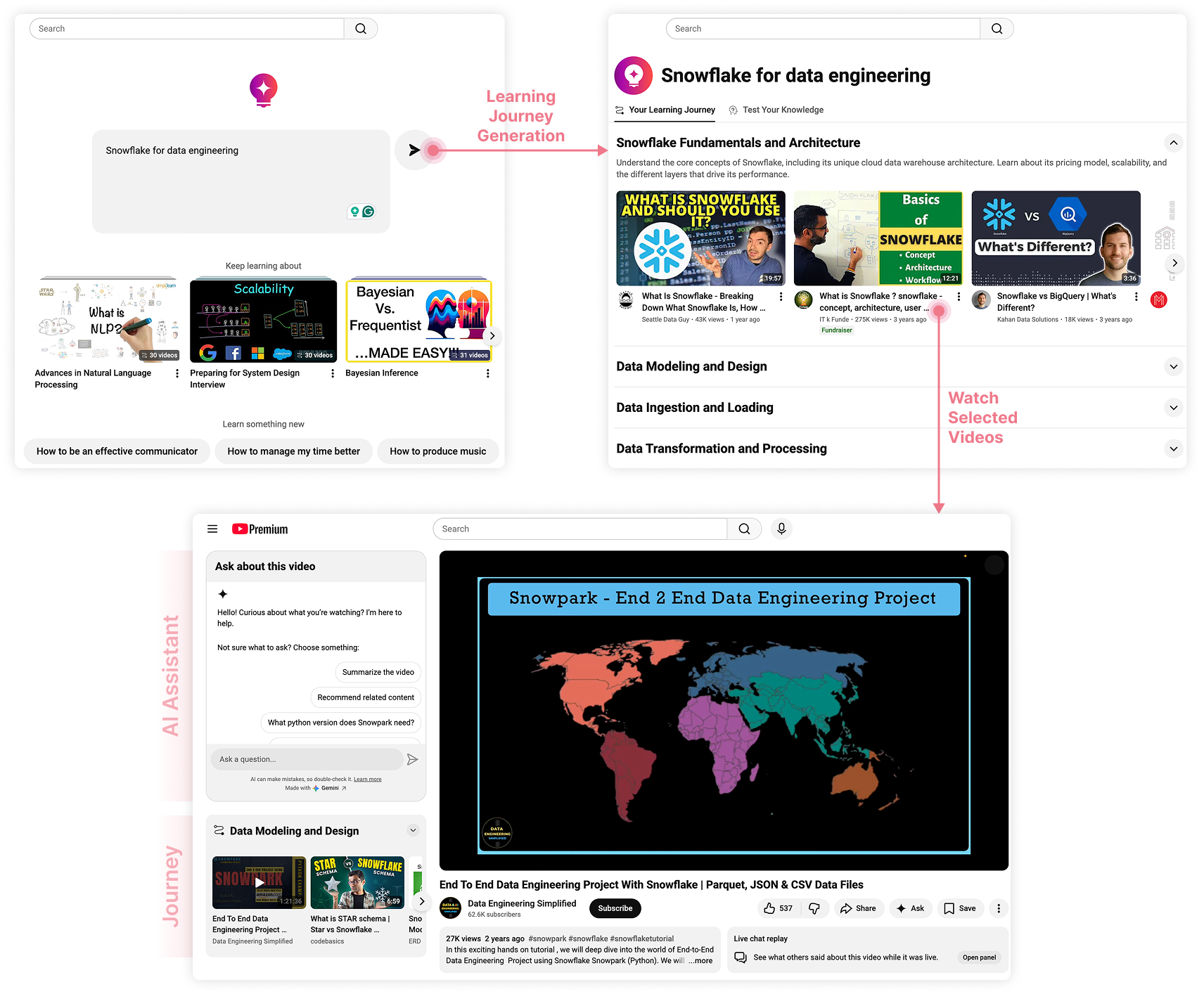}
  \caption{Overview of YouTube Learning workflow. A user enters a query to generate topic-based video groupings organized into categories. Learners browse and select videos from these grouped sections and watch them in the video interface, where a single-video AI assistant provides support during viewing.}
  \Description{Overview of YouTube Learning workflow. The top-left shows a search interface where a user enters a topic. The system generates a structured learning journey composed of topic-based sections with grouped videos. The top-right panel shows categorized video lists such as fundamentals, data modeling, and data ingestion. The bottom panel shows the video watching interface, where a selected video is played alongside a sidebar AI assistant that supports question answering and navigation. Arrows indicate the flow from query to learning journey generation to video consumption with assistant support.}
  \label{fig:youtube}
\end{figure*}

\section{Formative Study Participants}

\begin{table*}[t]
\centering
\caption{Participants background information and selected learning topics.}
\label{tab:formative}
\begin{tabular}{lllll}
\toprule
\textbf{Participant} & \textbf{Field} & \textbf{Age} & \textbf{Learning Topic} & \textbf{Stemmed From} \\
\midrule
P1 & Information Science & 25-34 & Mind and Body Problem & Course \\
P2 & Computer Science & 25-34 & Learning about how React works & Research \\
P3 & Computer Science & 25-34 & How to take care of house plants & Personal \\
P4 & Computer Science & 25-34 & Hawking Radiation Theory / Black Holes & Personal \\
P5 & Computer Science & 25-34 & Jazz Reharmonization & Course \\
\bottomrule
\end{tabular}
\end{table*}

\section{Formative Study Interview Questions}
\label{app:interview-questions}

\subsection{Setting Goals \& Skills Development}
\begin{itemize}
\item Do you usually have a learning goal before you start learning?
\item Do you feel you always know what you want to learn and why? Or do you wish you had more guidance?
\item How do you see informal continuous learning as a way to help you in systematic skills building?
\end{itemize}

\subsection{Knowledge Construction}
\begin{itemize}
\item How do you prefer to build your knowledge when you learn something new?
\item Do you usually like to explore many related topics (breadth), or do you prefer to go deeply into one area at a time (depth)?
\end{itemize}

\subsection{Supporting Learning Progression and Routines}
\begin{itemize}
\item When you go from one content to another in your learning journey, how do you track what you have learned?
\item Do you usually feel you are in control of the knowledge you build or where you left off?
\item What feels most confusing or frustrating about trying to use these platforms for long-term learning rather than one-off questions?
\item Why do you think using informal continuous learning would help? What learning outcomes / experiences can it bring to you?
\end{itemize}

\subsection{Future Design}
\begin{itemize}
\item What are some features or support that you think would make it much easier for you to start and maintain continuous learning in a topic?
\end{itemize}

\section{Pre-Survey Questions}
\label{app:pre-survey}

\subsubsection*{Section 1: Demographics}
\begin{enumerate}
    \item What is your age (range)?
    \item What is your gender?
    \item What is your highest level of education completed or currently pursuing?
    \item What is your current field of study or profession?
    \item Are you currently a student?
\end{enumerate}

\subsubsection*{Section 2: YouTube Learning Experience (7-point scale)}
\begin{enumerate}
    \item How often do you use YouTube for informal learning purposes?
    \item What kinds of topics do you typically learn on YouTube?
    \item How confident are you in organizing your own learning pathway on YouTube?
    \item When learning on YouTube, how often do you feel that videos logically build on each other?
    \item How often do you lose track of your learning progress on YouTube?
\end{enumerate}

\subsubsection*{Section 3: Experience with AI / LLM Tools (7-point scale)}
\begin{enumerate}
    \item How familiar are you with AI-powered tools (e.g., ChatGPT, Gemini)?
    \item How frequently do you use AI tools for informal learning?
    \item Have you used AI tools to generate study plans or structured learning pathways before?
\end{enumerate}

\section{Post-Survey Questions}
\label{app:post-survey}
Items marked with $\dagger$ were asked only for the YT-Pilot condition.

\subsubsection*{Phase 1A: Perceived Usefulness --- Goal-Setting \& Planning (5-point scale)}
\begin{enumerate}
    \item The system's input features enabled me to clearly define my personal learning goals.
    \item Using the system enabled me to reflect on my personal plan.
    \item[$\dagger$3.] The concept roadmap helped me see the big picture and plan my study around specific concepts.
\end{enumerate}

\subsubsection*{Phase 2B: User Experience --- Performance, Controllability \& Explainability (7-point scale)}
\begin{enumerate}
    \item The pathway was clearly presented.
    \item The pathway met my goals and needs.
    \item The videos in the pathway were logically connected with clear progression.
    \item I feel confident and able to revise the pathway easily to suit my specific goals and needs.
    \item The system helps me understand why certain videos were included in the pathway.
    \item[$\dagger$6.] The system clearly explains the connection between the weekly study tasks.
\end{enumerate}

\subsubsection*{Phase 3A: Perceived Usefulness, Perceived Ease of Use \& AI-Generated Notes (5-point scale)}
\begin{enumerate}
    \item Using this system improved my ability to learn new topics effectively.
    \item The AI-generated roadmap/pathway made it easier to structure and manage my learning.
    \item[$\dagger$3.] The context-based note-taking feature was useful for capturing and organizing key points while learning.
    \item Interacting with the AI assistant to get support while learning was helpful.
    \item Tracking my progress through the pathway required minimal effort.
    \item Overall, I found the system easy to use.
\end{enumerate}

\section{Participants Demographics}

\begin{table*}
\centering
\caption{Participant demographics ($N = 20$). YT Freq.\ = YouTube learning frequency (1--7 scale); AI Fam.\ = AI tool familiarity (1--7 scale).}
\label{tab:participant-demographics}
\small
\begin{tabular}{llllllcc}
\toprule
\textbf{PID} & \textbf{Age} & \textbf{Gender} & \textbf{Education} & \textbf{Field} & \textbf{Student} & \textbf{YT Freq.} & \textbf{AI Fam.} \\
\midrule
P1  & 25--34 & Female & Doctoral   & Computer Science       & Yes & 6 & 7 \\
P2  & 25--34 & Male   & Doctoral   & Computer Science       & Yes & 3 & 6 \\
P3  & 18--24 & Male   & Doctoral   & Computer Science       & Yes & 7 & 7 \\
P4  & 25--34 & Female & Master's   & Computer Science       & Yes & 2 & 6 \\
P5  & 25--34 & Male   & Master's   & Information Management & Yes & 3 & 7 \\
P6  & 25--34 & Male   & Master's   & Information Science    & Yes & 6 & 7 \\
P7  & 18--24 & Female & Master's   & Information Management & Yes & 6 & 6 \\
P8  & 25--34 & Male   & Master's   & Computer Engineering   & No  & 5 & 7 \\
P9  & 25--34 & Female & Doctoral   & HCI                    & Yes & 5 & 6 \\
P10 & 25--34 & Male   & Master's   & Information Management & Yes & 6 & 7 \\
P11 & 25--34 & Female & Master's   & Information Management & Yes & 2 & 5 \\
P12 & 18--24 & Male   & Bachelor's & Computer Engineering   & Yes & 6 & 7 \\
P13 & 25--34 & Male   & Doctoral   & Computer Engineering   & Yes & 3 & 7 \\
P14 & 18--24 & Male   & Master's   & Information Management & Yes & 7 & 6 \\
P15 & 25--34 & Male   & Doctoral   & Computer Science       & Yes & 3 & 6 \\
P16 & 25--34 & Female & Doctoral   & Computer Science       & Yes & 3 & 7 \\
P17 & 25--34 & Female & Master's   & Information Management & Yes & 4 & 7 \\
P18 & 25--34 & Female & Doctoral   & Informatics            & Yes & 5 & 6 \\
P19 & 25--34 & Male   & Doctoral   & HCI                    & Yes & 6 & 6 \\
P20 & 18--24 & Female & Bachelor's & Computer Science       & Yes & 2 & 6 \\
\midrule
\bottomrule
\end{tabular}
\end{table*}

\section{YT-Pilot System Prompts}
\label{app:system-prompts}

This appendix presents the main prompts used in each stage of the YT-Pilot system, as referenced in Section~4.
\subsection{Concept Map Generation Prompt}
\label{app:prompt-concept-map}

\begin{promptbox}{Concept Map Generation Prompt}
\syskw{System:} You are a curriculum designer. Return ONLY valid JSON, no markdown.

\userkw{Input:} Topic: "{{topic}}".

\keykw{Rules:} Return JSON exactly, include "description" as 1 sentence how this topic is structured, and "concepts" as a list of exactly {{numConcepts}} concepts, each with "label" as a 2-4 word name and "description" as 1 sentence what it covers. Use topic-specific labels, not generic like "Introduction". Order concepts from foundational to advanced (Bloom's Taxonomy).
\end{promptbox}

\subsection{Learning Pathway Generation Prompt}
\label{app:prompt-pathway}

\begin{promptbox}{Pathway Generation Ordering Prompt (Abridged)}
\syskw{System:} You are a curriculum designer. You are given REAL YouTube videos grouped by concept. Each concept represents one week of study.

\keykw{Bloom progression:} Week 1: Remember \& Understand (levels 1-2). Week 2: Understand \& Apply (levels 2-3). Week 3: Apply \& Analyze (levels 3-4). Week 4+: Analyze, Evaluate \& Create (levels 4-6). Bloom levels MUST increase across weeks. Within each week, videos also progress, slot 1 = lower, slot 3 = higher. The overall Bloom level should NEVER decrease.

\keykw{Critical rules:} Use content-based selection ranked by signal strength: transcript\_snippet (STRONGEST signal), chapters, tags, description, title, then quality metrics. The 3 videos in each week must form a coherent mini-sequence where each genuinely builds on the previous. The last video of week N must unlock a concept that the first video of week N+1 requires, and this must be a genuine prerequisite, not a vague topical connection. Explain what specific knowledge each week builds on and what it unlocks for the next. Never use the same video twice or select two videos that teach the same sub-topics.

\keykw{Output:} Return STRICT JSON with: course\_title, course\_description, bloom\_progression, learning\_objectives, weeks[{concept, focus, bloom\_levels, why\_this\_week\_first}], videos[{candidate\_index, bloom\_level, bloom\_verb, requires\_concept, unlocks\_concept, zpd\_rationale, learning\_objective, why\_selected, dependency\_explanation, keywords}].
\end{promptbox}

\subsection{AI Study Assistant Prompts}
\label{app:prompt-assistant}

\begin{promptbox}{Study Assistant --- System Prompt}
\syskw{System:} You are a personal tutor. Be direct and concise. By default use plain short sentences, but if the user asks for bullet points, lists, formatting, or any specific structure, follow their request. Answer confidently based on the information provided. Never say "likely", "maybe", "probably", or "I think". If you have video metadata but no transcript, use the title, concept, and description to give a definitive answer.
\end{promptbox}

\begin{promptbox}{Study Assistant --- Question Classification Prompt}
\syskw{System:} Classify this student question into one category, A) about the CURRENT video content, concepts, explanations, details from what they are watching, or B) about the PATHWAY, other videos, what to watch next, comparisons, progression, recommendations, and connections between videos.

\userkw{Input:} Question: "{{message}}".

\keykw{Output:} Respond with just A or B.
\end{promptbox}

\begin{promptbox}{Study Assistant --- Response Prompt Template}
\syskw{System:} Learning pathway: "{{topic}}". Currently watching: "{{video\_title}}" by {{instructor}}. Progress: {{completed}}/{{total}} videos completed. Current video transcript: {{transcript (up to 3,000 characters)}}. PATHWAY VIDEOS ({{N}} total): 1. "Video Title" (Concept) [CURRENT VIDEO], 2. "Video Title" (Concept), ...

\keykw{Rules:} IMPORTANT: When referencing videos, ONLY use exact titles from this list. Never invent or suggest videos outside this pathway. 

{{For Type B only: DETAILED VIDEO CONTEXT: 1. "Video Title" Concept: ... Desc: ... Transcript excerpt: ... (up to 400 chars) --- 2. "Video Title" ... }}

Conversation so far: {{last 6 messages}}. Student: {{message}}.
 {{Type A: "Tutor (reply concisely in 2-4 sentences):"}} {{Type B: "Tutor (reference specific videos from the pathway by their exact title):"}}
\end{promptbox}

\subsection{Context-Based Note-Taking Prompt}
\label{app:prompt-notetaking}

\begin{promptbox}{Context-Based Note-Taking --- With Transcript Segment}
\syskw{System:} You are a study note assistant. Generate notes based ONLY on the transcript content provided. Return 2-3 bullet points starting with bullet markers. Each bullet must reference specific ideas, terms, or examples from the transcript. Never repeat ideas from previous notes. No headers, no markdown.

\userkw{Input:} Video: "{{title}}". Topic: {{main\_concept}}. Key terms: {{keywords}}. Learning goal: {{learning\_objective}}. Transcript around {{timestamp}}: "{{transcript segment (plus or minus 60-second window)}}". Student paused at {{timestamp}}.

Previous notes: [{{timestamp}}] {{note content}} ...

Write NEW points about what is being discussed at {{timestamp}} that are DIFFERENT from the notes above.
\end{promptbox}

\begin{promptbox}{Context-Based Note-Taking --- Without Transcript (Fallback)}
\syskw{System:} You are a study note assistant. Generate notes about what is being taught at this point in the video. Return 2-3 bullet points starting with bullet markers. Each bullet should be one concise sentence. Never repeat ideas from previous notes. No headers, no markdown.
\end{promptbox}

\section{System Logic}

\begin{figure*}
  \centering
  \includegraphics[width=\textwidth]{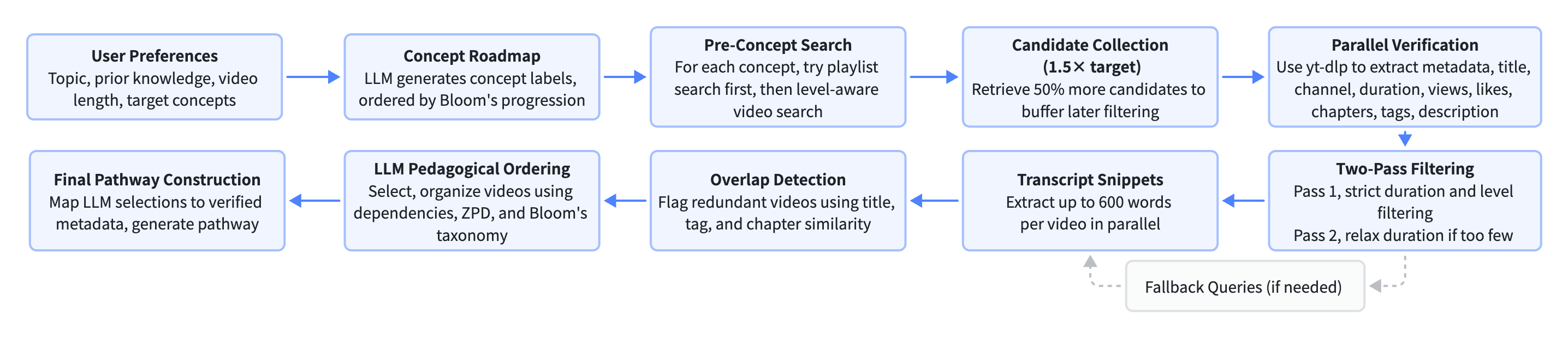}
  \caption{Pathway generation pipeline in YT-Pilot. Starting from user preferences, the system constructs a concept roadmap, retrieves and verifies candidate videos, extracts transcripts, and filters content. An LLM then performs pedagogical ordering using conceptual dependencies, ZPD, and Bloom’s taxonomy to produce a structured multi-video learning pathway.}
  \Description{Pipeline diagram illustrating pathway generation. The process begins with user preferences including topic, prior knowledge, and constraints. An LLM generates a concept roadmap organized by Bloom's taxonomy. For each concept, the system performs playlist-first and video-level search, collects candidate videos, and verifies metadata using yt-dlp. Transcript snippets are extracted in parallel. A two-pass filtering process selects videos based on duration and level, with fallback queries if needed. Overlap detection removes redundant content. Finally, an LLM performs pedagogical ordering using dependencies, Zone of Proximal Development, and Bloom's taxonomy to construct a structured learning pathway.}
  \label{fig:pathway_logic}
\end{figure*}

\begin{figure*}
  \centering
  \includegraphics[width=0.78\textwidth]{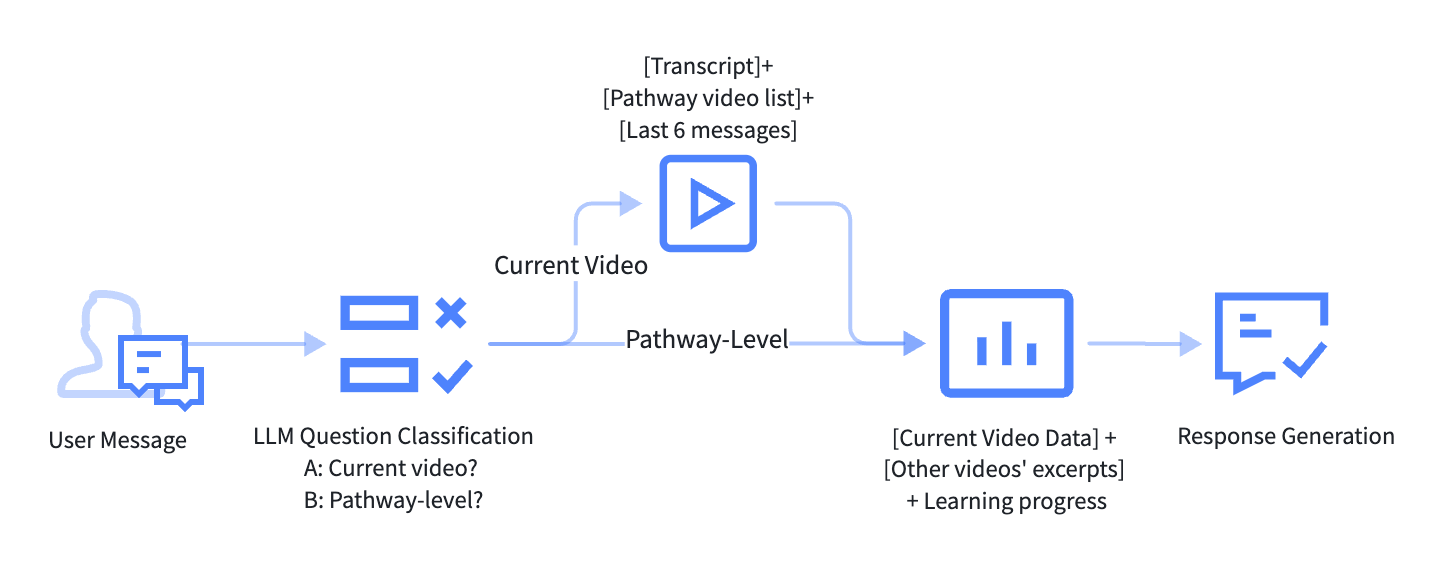}
  \caption{Patyway-aware assistant for question classification and context assembly. User queries are classified as current-video or pathway-level, and context is assembled from either the active video or across multiple videos using transcripts, pathway structure, and learning progress to generate responses.}
  \Description{Diagram of the AI study assistant workflow. A user message is first classified by an LLM into either a current-video question or a pathway-level question. For current-video questions, the system uses transcript and recent interaction context from the active video. For pathway-level questions, it aggregates information from multiple videos, including transcripts, pathway structure, and recent messages. The assembled context is then passed to a response generation module, which produces an answer grounded in either local or cross-video context.}
  \label{fig:aiassistant_logic}
\end{figure*}

\begin{figure*}
  \centering
  \includegraphics[width=0.78\textwidth]{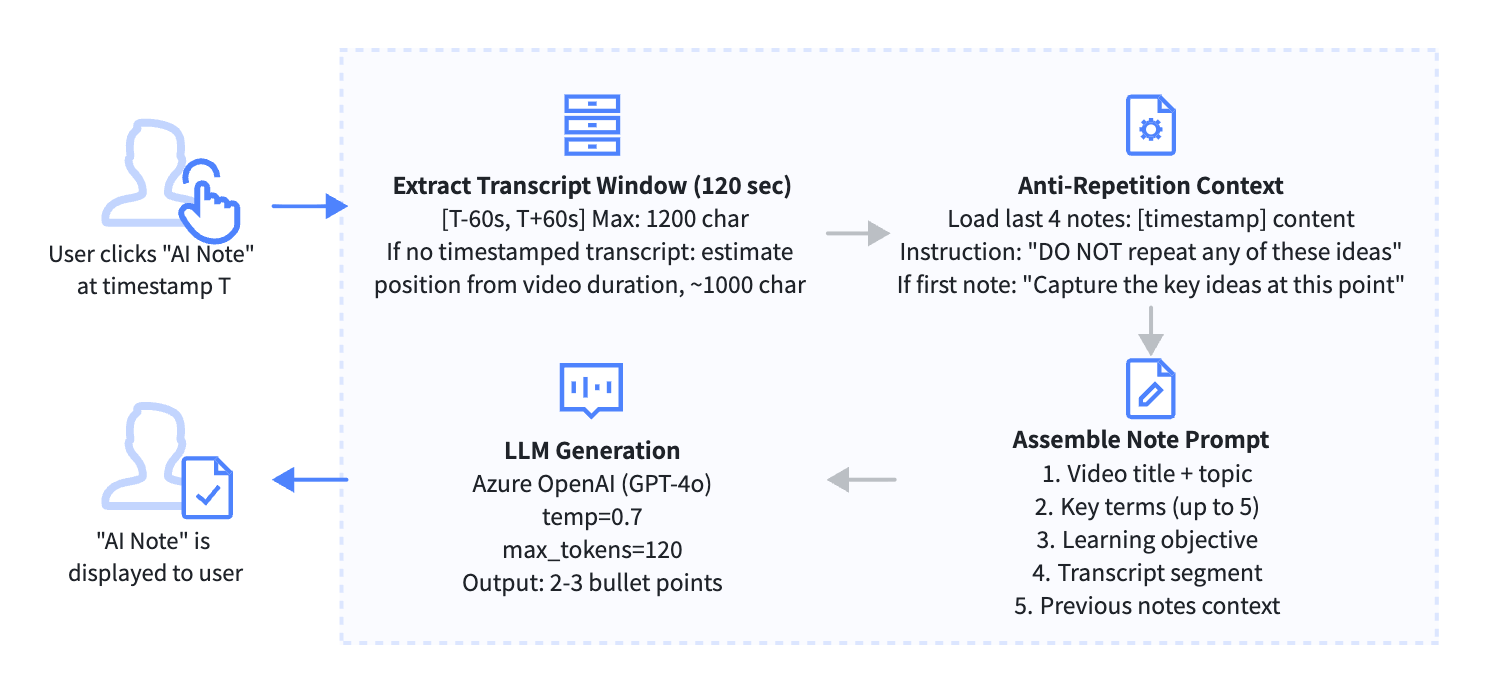}
  \caption{Context-based note-taking with timestamp extraction and anti-repetition. The system captures transcript context around the current timestamp, incorporates prior notes and learning context, and generates concise notes while avoiding redundancy, anchoring them within the learning pathway.}
  \Description{Diagram of context-based note-taking process. When a user clicks the AI note button at a specific timestamp, the system extracts a transcript window around that time. It loads recent notes to avoid repetition and constructs a prompt including video metadata, key terms, learning objectives, and transcript content. The LLM generates concise bullet-point notes. An anti-repetition mechanism ensures new notes do not duplicate previous content. The generated note is then displayed to the user and anchored to the corresponding timestamp within the learning pathway.}
  \label{fig:notetaking_logic}
\end{figure*}

\clearpage 
\bibliographystyle{ACM-Reference-Format}
\bibliography{references} 

\end{document}